# Comparison between upconversion response of $Er^{3+}$ sensitised with $Yb^{3+}$ in various oxidic ceramic hosts

**Liviu Dudaș*, Daniela Berger and Cristian Matei**

Faculty of Chemical Engineering and Biotechnologies, National University of Science and Technology Politehnica Bucharest, 1-7 Gheorghe Polizu Street, 011061 Bucharest, Romania; liviu_dudas@yahoo.com (L.D.); daniela.berger@upb.ro (D.B.), cristian.matei@upb.ro, (C.M.)
* Correspondence: liviu_dudas@yahoo.com (L.D.);

**Abstract**: The upconversion process for the $Er^{3+}$ ion, when irradiated with IR photons at 980 nm, strongly depends upon the presence of the sensitizer $Yb^{3+}$ ions. There are many studies analyzing the properties of the upconversion process for various crystalline ceramic matrices, but these, in their overwhelming majority, focus on only one compound. Comparative studies are very scarce, and the treatment of each case is limited. The purpose of this study is the comparison of the upconversion response of $Er^{3+}$ in some $Er^{3+}$:$Yb^{3+}$-doped oxidic ceramics. This comparison helps to observe aspects of the phenomena that are common across the cases, like the variation of red-green intensity ratios when the sensitizer's $Yb^{3+}$ concentration is increased, offering hints about the mechanism by which $Er^{3+}$ is sensitized by $Yb^{3+}$. Sol-gel methods were used to obtain doped oxidic ceramics, which were characterized by XRD and SEM and their upconversion spectra were measured. There is a good correlation between the relative and absolute concentrations of the activator and sensitizer species and the intensities of the emission lines in the visible spectra. We observed that the ratios between emission intensities in the green band (510–580 nm) and red band (640–700 nm) (i.e., the spectral content) show similarities between different host crystals for $Er^{3+}$ and $Yb^{3+}$, which are indications that, regardless of the crystalline medium of hosting, the dopant ions interact in some specific and similar ways, and a hypothesis explaining this is suggested.

## 1. Introduction

Lanthanide incident radiation upconversion (UC) consists of an ion absorbing energy, either photonic or from the lattice of embedding, through successive steps, transitioning to higher energy states [1] of its metastable $4f^n$ electron configurations.When the ion reaches a certain state, which has a higher energy and shorter lifetime than the previous ones, it de-excites and emits a photon that has an energy greater than each of the previous absorbed ones. [2]

This phenomenon is the opposite of downcoversion (DC), when a single high-energy photon excites the ion on a high energy level from where it de-excites to ground state through successive steps, passing through the intermediary states and emitting multiple photons with lower energies than that of the absorbed one.

This process is sensitive to the host of the lanthanide ions; each gaseous, liquid, amorphous solid, or crystalline solid environment has its own kind of influence upon it [3]. The upconversion is used in many domains like luminescent phosphors [4], nanoscale sensitive thermometry [5, 6], light-triggered drug delivery in targeted tissues [7] etc. and following is some information about the compounds studied in this work.

Yang, Liu et al., in [8], in order to increase silicon-based solar cells, study $BaGd_2ZnO_5$ (**BGZ**) doped with $D^{3+}$ and $Yb^{3+}$ and search for the NIR quantum cutting-down conversion luminescence, also finding three photon processes under 970 nm illumination and finding quantum efficiency up to 158%. Zhou & Zhang, in [9], synthesize BGZ doped with $Er^{3+}$:$Yb^{3+}$ by solid-state reaction (SSR) and observe the two photon processes for green and red emissions, proposing a model for explaining the temperature quenching of green emissions. Yang, Liu et al., in [10], also trying to improve solar cells efficiency, are preparing BGZ:Er:Yb by the sol-gel method (SGM) and find notable luminescence for Ce:Yb dopants. Sun, Liu et al., in [11], are using the sol-gel technique to synthesize BGZ:Tm:Yb for temperature sensing and, for 0.1%Tm:12%Yb, find strong UC emissions at blue (478 and 485 nm) when illuminated with 980 nm, and the involved three-photon process is discussed. Georgescu et al., in [12], study the phonon sidebands of $Eu^{3+}$ when doping BGZ, find an average energy for phonons at 318 $cm^{-1}$ and remark that the values of the electron-phonon coupling strengths are comparable with those in other oxidic materials; they also calculate the probabilities for multiphonon transitions.

Xie, Mei et al., in [13], are synthesizing $BaLa_2ZnO_5$:Tm:Yb (**BLZ**) using the high temperature solid-method, finding efficient UC emissions at 478 nm and noting the two and three-photon processes for the weaker red emissions. Xie, Mei et al., in [14], synthesizing BLZ:Ho:Yb, find the predominant luminescence at 548 nm and find the optimal doping at 0.75% Ho and 15% Yb and remark that the combination is good for yellow-green UC phosphors. Singh et al., [15], synthesize, by SGM, BLZ:Gd activated phosphors for UV-B, which have narrow band peak emission at 314 nm. Also, Singh et al., in [16], are using the SGM for BLZ:Tb phosphors, noting the shift of emission from blue to green when increasing Tb concentration, and propose the use of these phosphors in white LEDs because they are converting UV to green. Singh et al., in [17], obtain BLZ:Sm red-orange (603 nm) emitting phosphors when excited with blue at 410 nm, and note the concentration quenching.

Sonika et al., in [18], synthesize by solution-combustion method (SCM) $BaY_2ZnO_5$ (**BYZ**) doped with Dy nanophosphors, record photoluminescence spectra upon excitation with 355 nm, note the 580 nm emission, and also note its quenching when Dy concentration is at 4%. In [19], Chara et al. are using urea-assisted, single-step SCM to synthesize BYZ doped with Sm, excite the spherical nanoparticles with 411 nm, and note the 610 nm emission suitable for white LED. Also, concentration quenching is seen beyond 3% due to dipole-dipole interactions. Li, Wei et al., in [20], are obtaining BYZ:Eu by SSR at high temperature, measuring fluorescence decay times, and using 469 nm for excitation to probe the $^5D_0$ level of $Eu^{3+}$ at various temperatures and show that this phosphor is a good candidate for luminescence thermometry in the range of 330 - 510 K. BYZ:Eu is also studied by Dalal et al., in [21], who use the glycine-assisted SCM and note the color tunability of these nanophosphors that could be used for efficient white LEDs at near-UV excitation. Chahar et al., in [22], are also synthesizing BYZ:Er by the urea-assisted SCM and find the 549 green emission of Er when excited with 380 nm. Shih, in [23], prepares BYZ:Tm by vibrating mill SSR and measures, at illumination with 362 nm, a maximum 458 nm emission for 2% $Tm^{3+}$, with color purity better than the commercial blue phosphors. In [24], Fan, Liu et al. produced BYZ:Dy:Sm by high-temperature SSM, excites it with 351 nm, and finds an energy transfer efficiency of 14% and adjusts the emission color by varying the Sm concentration. Very good for near-UV-excited white phosphpors.

Xu & Liu, in [25], created UC films, by using SGM and spin coating, of $Y_2O_3$ (**YO**) doped with Er:Yb for study of the $Li^+$ ions impact on the energy level populations of $Er^{3+}$. The crystal grains were enlarged, and the lattice symmetry was hindered. They reported a significant reduction of the non-radiative transitions and quenching and an enhanced 556 nm emission and also a shift in blue emission from 436 nm to 409 nm when Er:Li concentrations exceeded 1%:2% and considered $Li^+$ ions as changers of the YO matrix phonon energies. Liu & Liu, in [26], synthesized YO:Er:Gd by SGM and noted that $Gd^{3+}$ enhanced green emission of $Er^{3+}$ and used this compound as remote temperature-sensing phosphor. In [27], Tadge, Yadav et al., studied YO:Tm:Ho:Yb and changed the color point by adjusting $Yb^{3+}$ concentration and noting the shift toward red of the UC emission. They incorporated these nanophosphors into dye-sensitive solar cells and increased their power efficiency by over 10%. Zhao & Wu, in [28], used 1.5 μm laser to generate UV UC emission of YO:Yb:Er and concluded that $Er^{3+}$ acted as both as activator and sensitizer, suggesting a new method of tuning UC radiation for RE-doped materials. Yang, Liu et al. [29] also incorporated Li into YO:Er:Yb powders synthesized by the Pechini method and noted how this fact increases the UC emission intensity, assuming that $Li^+$ ions reduce the number of defect centers and quenching clusters.

As seen, these studies focus on the best method of synthesis and obtaining the best efficiencies for UC or DC, yet they do not make comparisons of the behavior of certain RE ions as dopants in different hosts. This article does not only present the comparative assessment of the Er:Yb activator sensitizer pair across the compounds aforementioned but also observes the common behavior, across all hosts, of the correlation between the increasing of the red emission intensity of $Er^{3+}$ when $Yb^{3+}$ concentrations increase—a fact that is not mentioned in literature and, as such, not explained.

## 2. Materials and Methods

**Table 1** presents the oxides used as crystalline ceramic hosts for $Er^{3+}$ and $Yb^{3+}$, the labels for the matrices, and the concentrations of $Er^{3+}$ and $Yb^{3+}$.

The relative concentration ratios of dopants were multiples of 2. These ratio choices were made because the results would be easier to correlate with the input parameters and would be easier to reveal what kind of dependence law governs the energy transfers between activators and sensitizers,: a logarithmic law or a linear one.

**Table 1**. The chemical formula of the oxide, the labeling of the ceramics made from it and the concentrations of $Er^{3+}$ and $Yb^{3+}$ which were used

| Host matrix | Label | %$Er^{3+}$ - %$Yb^{3+}$ |
|---|---|---|
| $Y_2O_3$ | YO | 1-0, 1-2, 1-4, 1-8 |
| $BaGd_2ZnO_5$ | BGZ | 1-0, 1-2, 1-4, 1-8, 2-4, 3-7 |
| $BaLa_2ZnO_5$ | BLZ | 1-2, 1-4 |
| $BaY_2ZnO_5$ | BYZ | 1-2, 1-4 |
| $Y_2TiO_5$ | YTO | 0-0, 0-1, 1-0, 1-2, 1-4, 1-8, 2-4, 4-4, 3-6 |

### *2.1. Synthesis of the oxide ceramics*

The oxide powders were synthesized by either citrate-EDTA method in aqueous medium or Pechini technique using ethylene glycol (EG)-based sol-gel (citrate or citrate EDTA) methods [30], in which the precursor metal salts, either nitrates, citrates, or acetates, in the desired stoichiometric ratios, were mixed.

The gels obtained, verified to be totally transparent, any trace of turbidity being a sign of error, were first calcined at 400 °C until the organic residues were charred, then further kept at 900 °C for 3 hours, after which white (pinkish) mixtures of oxides were obtained.

These powders were pressed into pellets, which were sintered at the appropriate temperatures (1100 °C to 1300 °C) and durations (3 h to 16 h) according to the type of ceramic.

The ceramic pellets were, in most cases, immersed into the oxidic powder of provenance when placed into the alumina crucibles in order to minimize the contamination, evaporation of the constituents, or any composition change at the surface.

### *2.2. X-ray diffraction*

The intermediate oxidic powders were investigated by X-ray diffractometry to check if the precursor phases are homogeneous. The eventual non-homogeneities should be fixed before sintering, because the thermal treatment, even if at high temperatures, does not guarantee the good mixing of the initial different phases and the formation, at the end, of the single desired crystalline phase.

The measurements were performed with a Rigaku Miniflex II X-ray diffractometer (Tokyo, Japan), using Cu-K$\alpha_1$ line, in a range of $2\theta$ of 15°–70°, 0.01° step and seed 2 °/min.

Once sintered, each ceramic pellet was again verified that corresponds to the desired compound, either by comparing the XRD with the simulated one for the theoretical structure, or with the appropriate ICDD cards. Figures 2-6 present visualizations of the crystal structures of the respective unit cells (CIF data from [31], [32] for visualizing), and the X-ray diffractograms, measured and compared [33], for the respective case.

### *2.3. SEM investigations*

In all cases, SEM investigations were performed for the resulted ceramic pellets and oxidic powders, which were not pressed into pellets but were submitted to the same sintering conditions as the pellets because they were acting as environmental coatings during sintering for these.

The SEM microscope used was a Tescan Vega 3LM equipped with an EDS spectrometer (Brno, Czech Republic).

Not only we observed the phase changing during sintering; some nanoparticles were melting (YO, BGZ), and some were just becoming glassy (YTO), but also, through EDAX, the eventual modifications of the compositions, either loss of constituents through evaporation or segregations, were checked.

### *2.4. Emission spectra measurement*

The spectra measuring of the samples was done with an Ocean Optics USB4000CG-UV-NIR spectrometer (Orlando, Florida, USA). The acquisition software was OceanView v. 1.6.7.

The measurement setup is described in [34]. This setup is peculiar to our methods and allows us to focus the laser beam on different spots on the pellet and check the local upconversion emission rather than an emission on a wider surface, which, usually, consists of a mixture of different spectral compositions.

In **Figure 1** is depicted the shape of the laser line that targets the samples, showing the level of accuracy of the exciting wavelength.

The spectra measured were for continuous-wave only. Even if this procedure ignores the fluorescence times [35], it still gives sufficient information in order to grasp a good understanding of the energy transfers between activators and sensitizers.

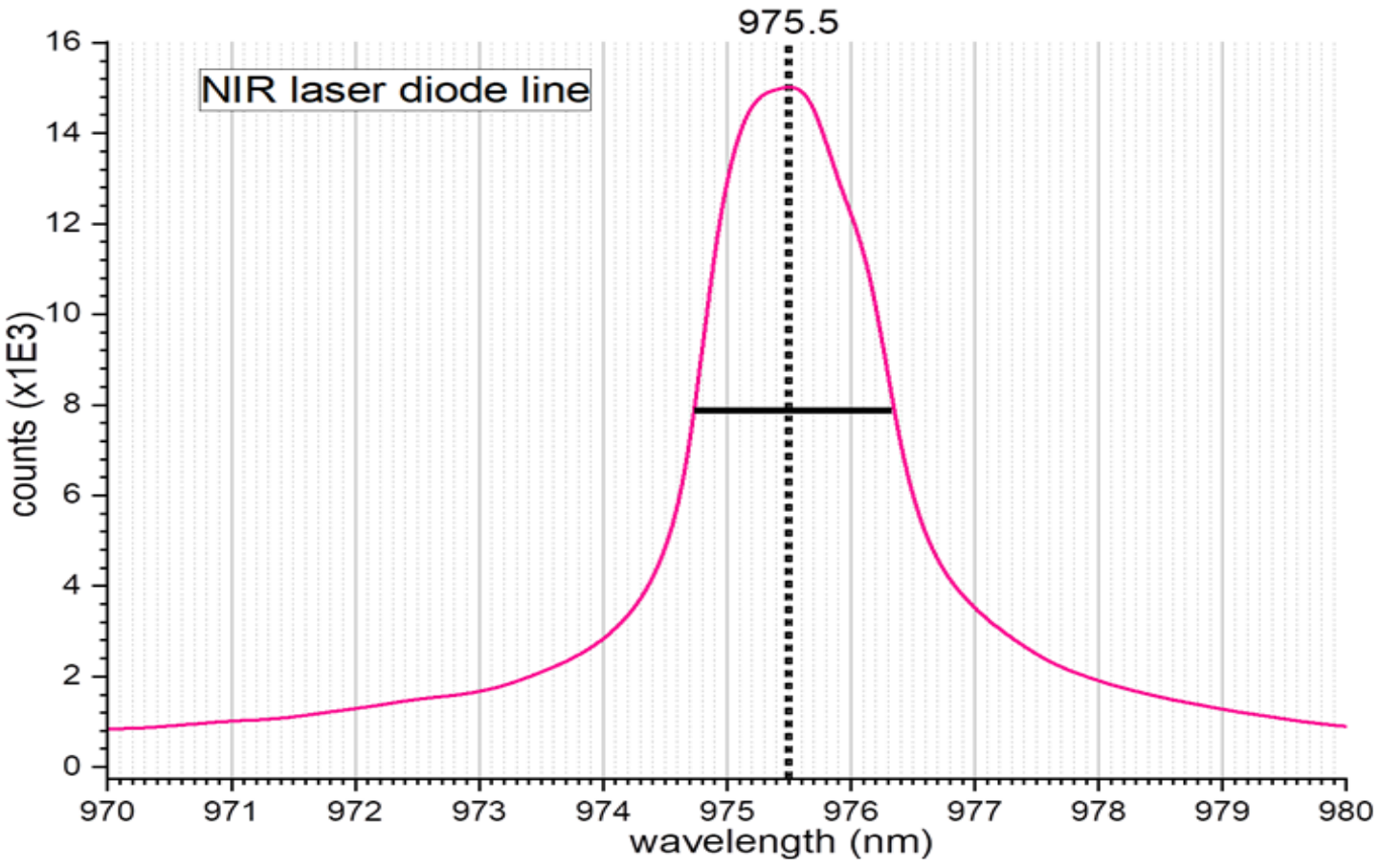

**Figure 1.** The emission line of the laser diode. The peak is at 975.5 nm and the FWHM is 2 nm.

## 3. Results

### *3.1 XRD characterization of ceramic pellets*

**Figures 2-6** present, in the left, the surrounding polyhedra of the $Y^{3+}$, $Gd^{3+}$ or $La^{3+}$ ions which are substituded by the dopant $Er^{3+}$ and $Yb^{3+}$ ions and at right, the X-ray diffractograms of the synthesized ceramics.

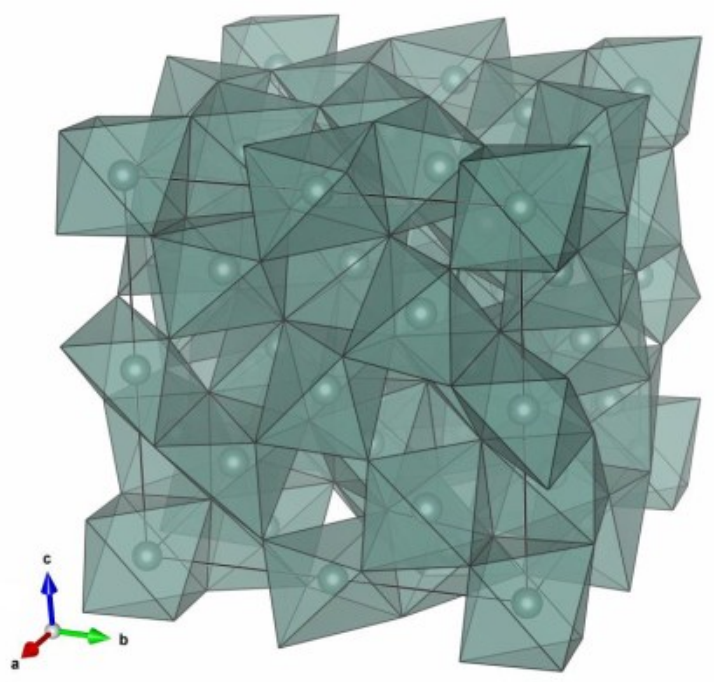

(A)

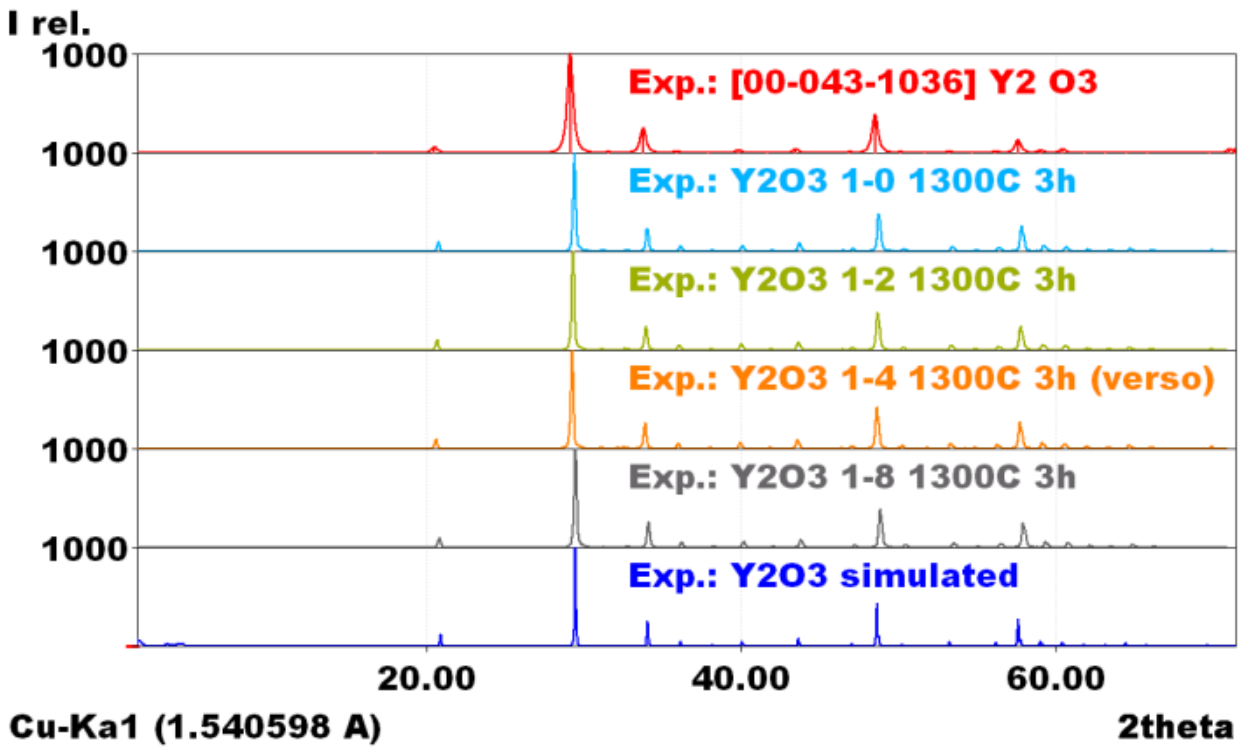

(B)

**Figure 2.** (A) Crystalline structure of $Y_2O_3$ matrices; (B) X-Ray diffraction patterns for YO pellets. The match between the simulated results and the measured data is very good.

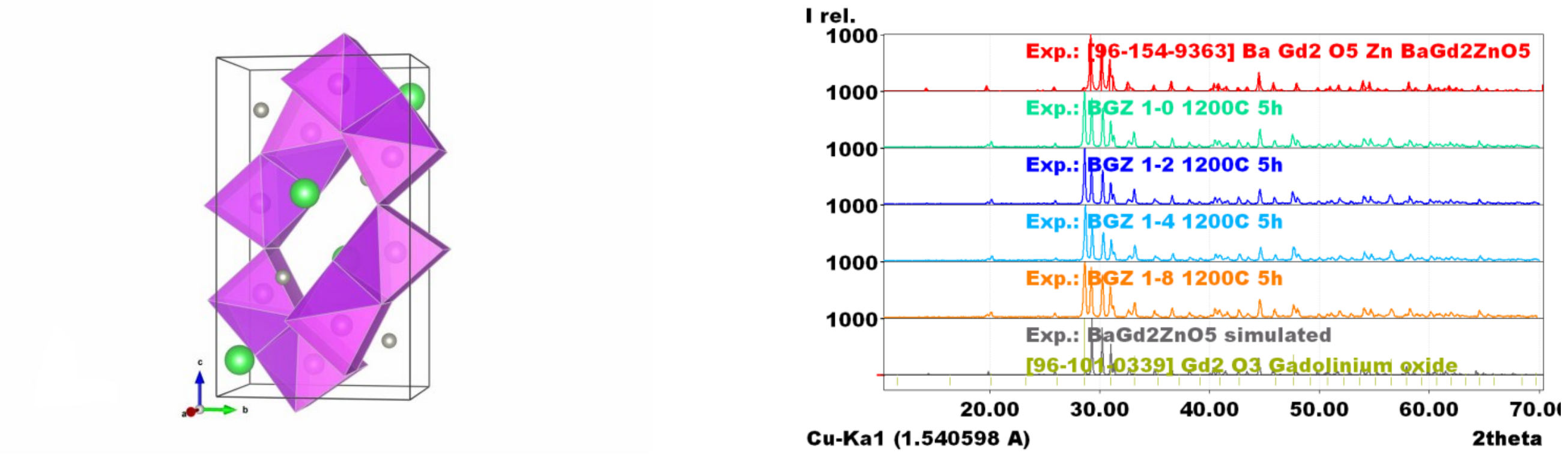

(A) (B)

**Figure 3.** (A) Crystalline structure of $BaGd_2ZnO_5$ matrices; (B) X-Ray diffraction patterns for BGZ pellets. The measured diffractograms show a combination of well-developed two phases, $BaGd_2ZnO_5$ and an excess of $Gd_2O_3$

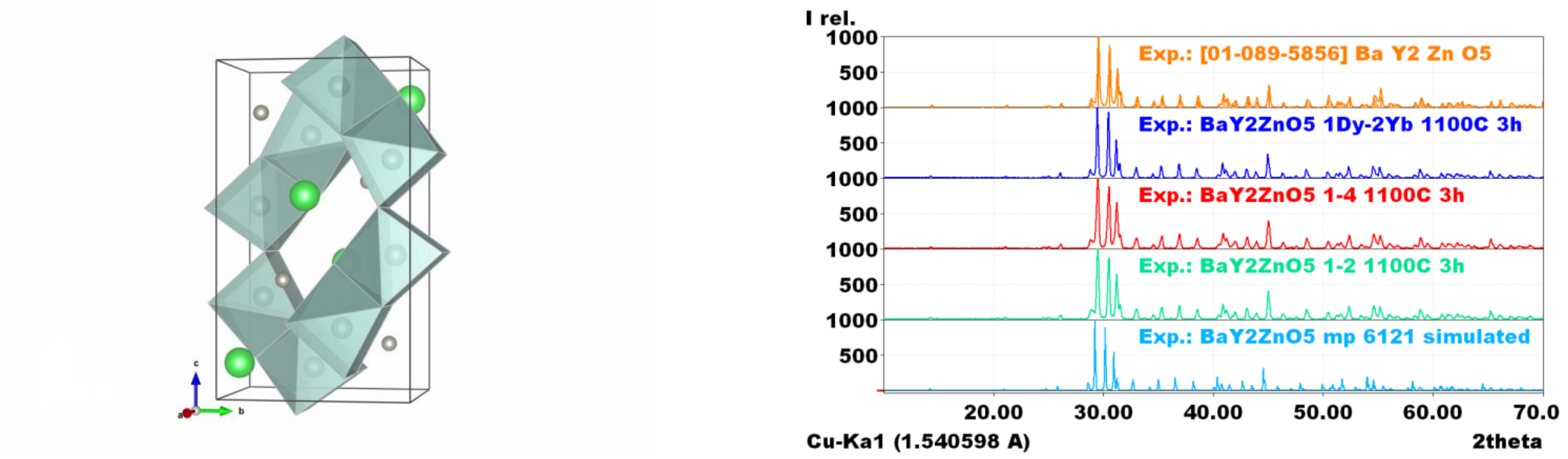

(A) (B)

**Figure 4.** (A) Crystalline structure of $BaY_2ZnO_5$ matrices; (B) X-Ray diffraction patterns for BYZ pellets. The match between the simulated results and the measured data is very good.

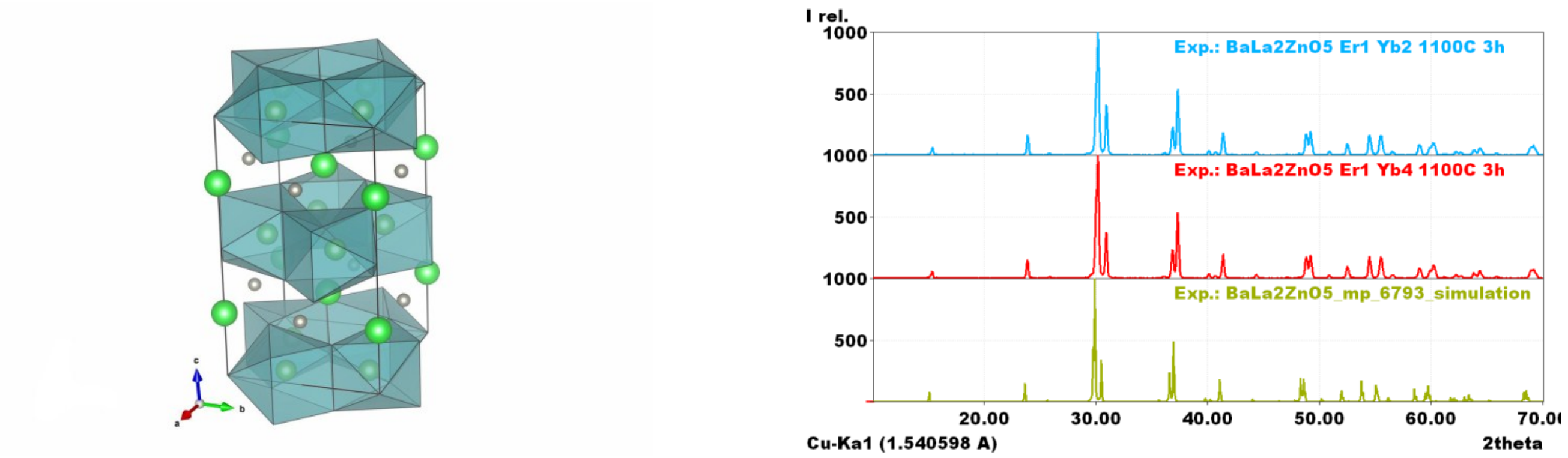

(A) (B)

**Figure 5.** (A) Crystalline structure of $BaLa_2ZnO_5$ matrices; (B) X-Ray diffraction patterns for BLZ pellets. The match between the simulated results and the measured data is very good.

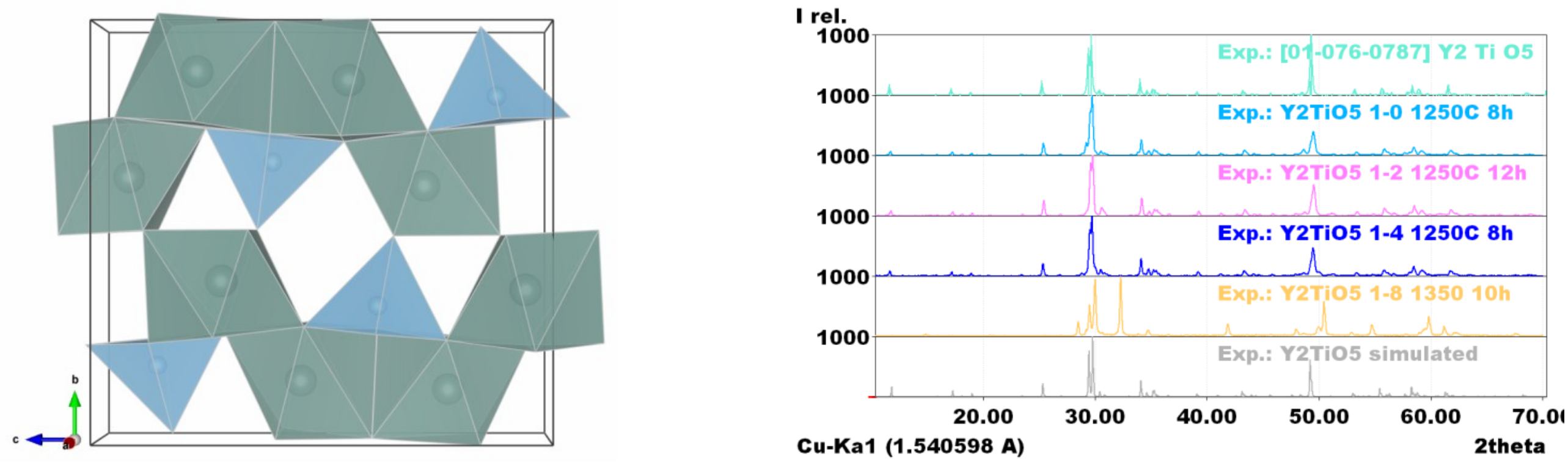

(A) (B)

**Figure 6.** (A) Crystalline structure of $Y_2TiO_5$ matrices; (B) X-Ray diffraction patterns for YTO pellets. The match between the simulated results and the measured data is very good, only for YTO 1-8 there is a strong deformation of the structure due to lower ionic radius of $Yb^{3+}$ vs the substituted $Y^{3+}$.

## *3.2. SEM investigation of the ceramic pellets*

In **Figures 7-11** are selected some typical SEM images of samples, usually 1-2 and 1-4 for the sintered ceramics. Observe how in the case of BGZ and YO, the sintering treatment lead to the melting and mixing of the precursor oxidic nanoparticle, while in the case of BLZ or YTO, the aggregation is only superficial. Maybe, in the case of BGZ, the temperature was too high and the $Gd_2O_3$ phase segregated due to lower energy of formation.

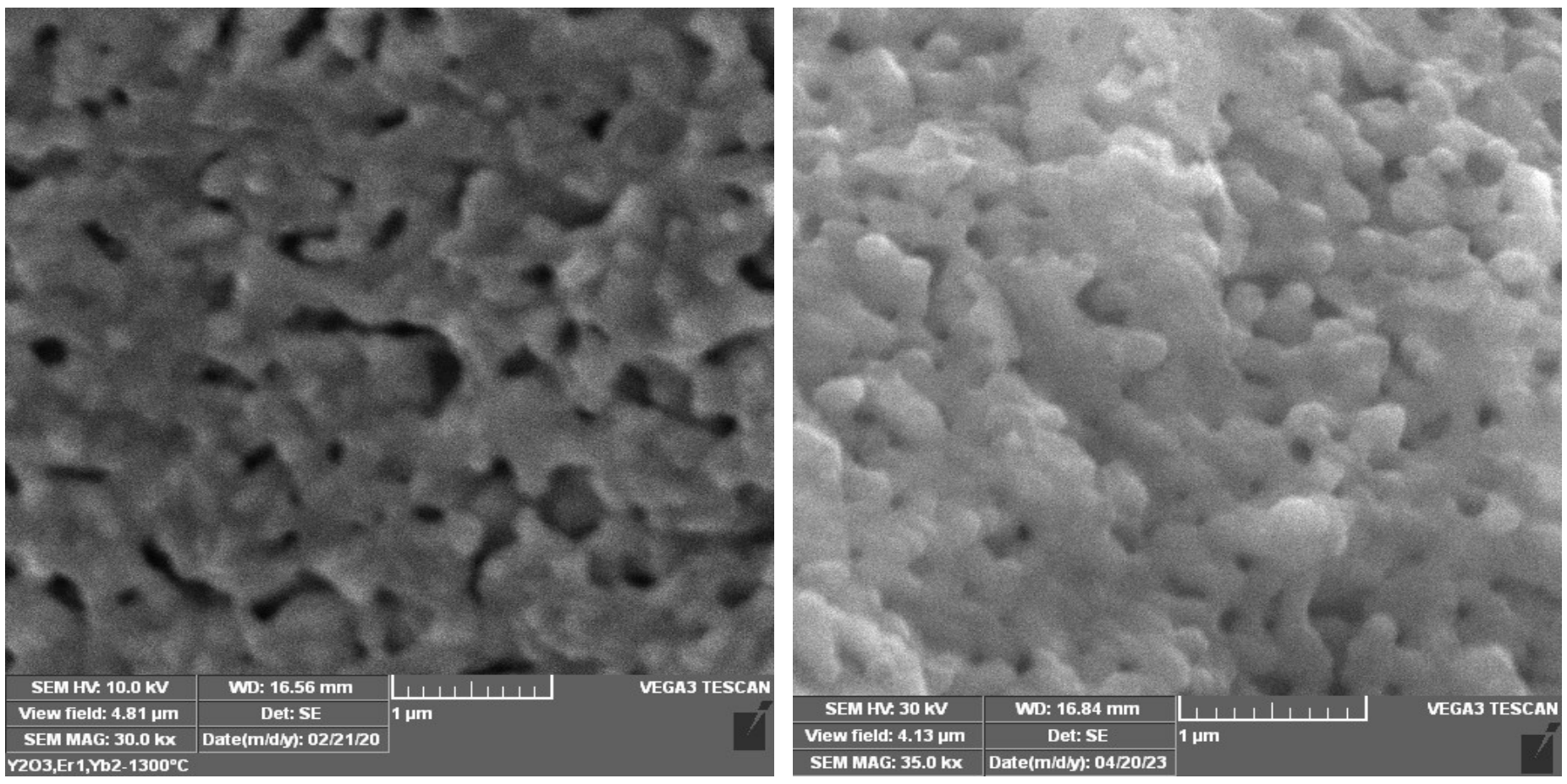

**Figure 7.** SEM for $Y_2O_3$ ceramics: (left) YO 1-2 and (right) YO 1-4 pellets.

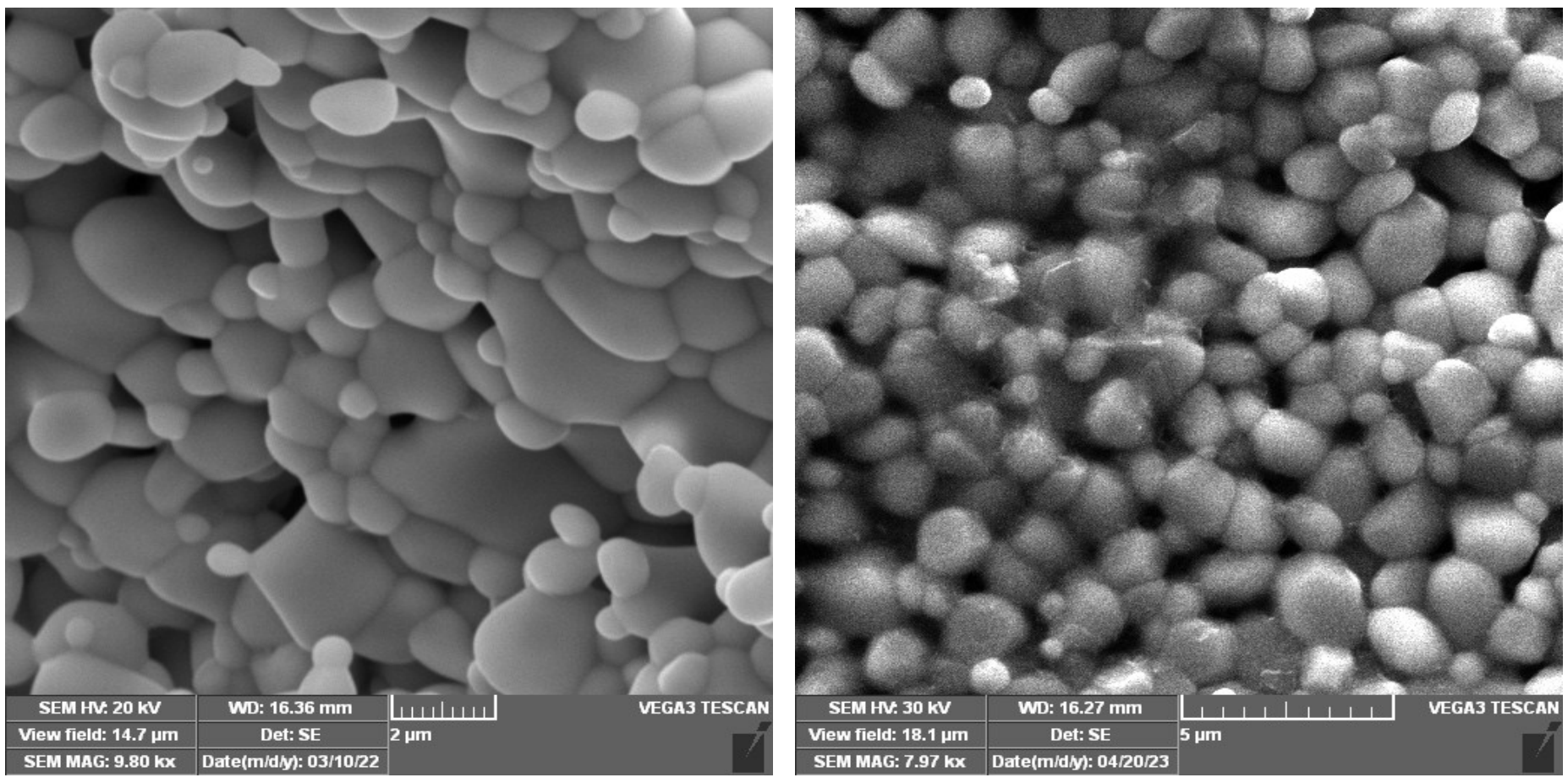

**Figure 8.** SEM for $BaGd_2ZnO_5$ ceramics: (left) BGZ 1-2 and (right) BGZ 1-4 pellets.

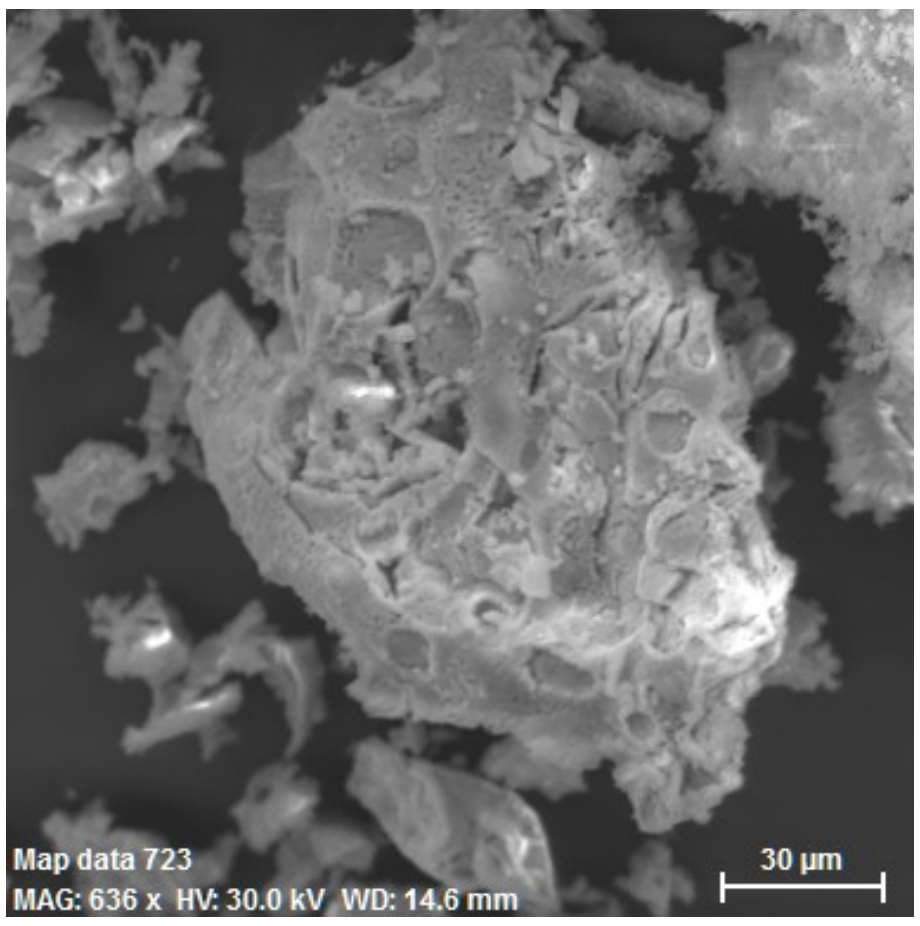

**Figure 9.** SEM for $BaLa_2ZnO_5$ BLZ 1-2 pellet interior.

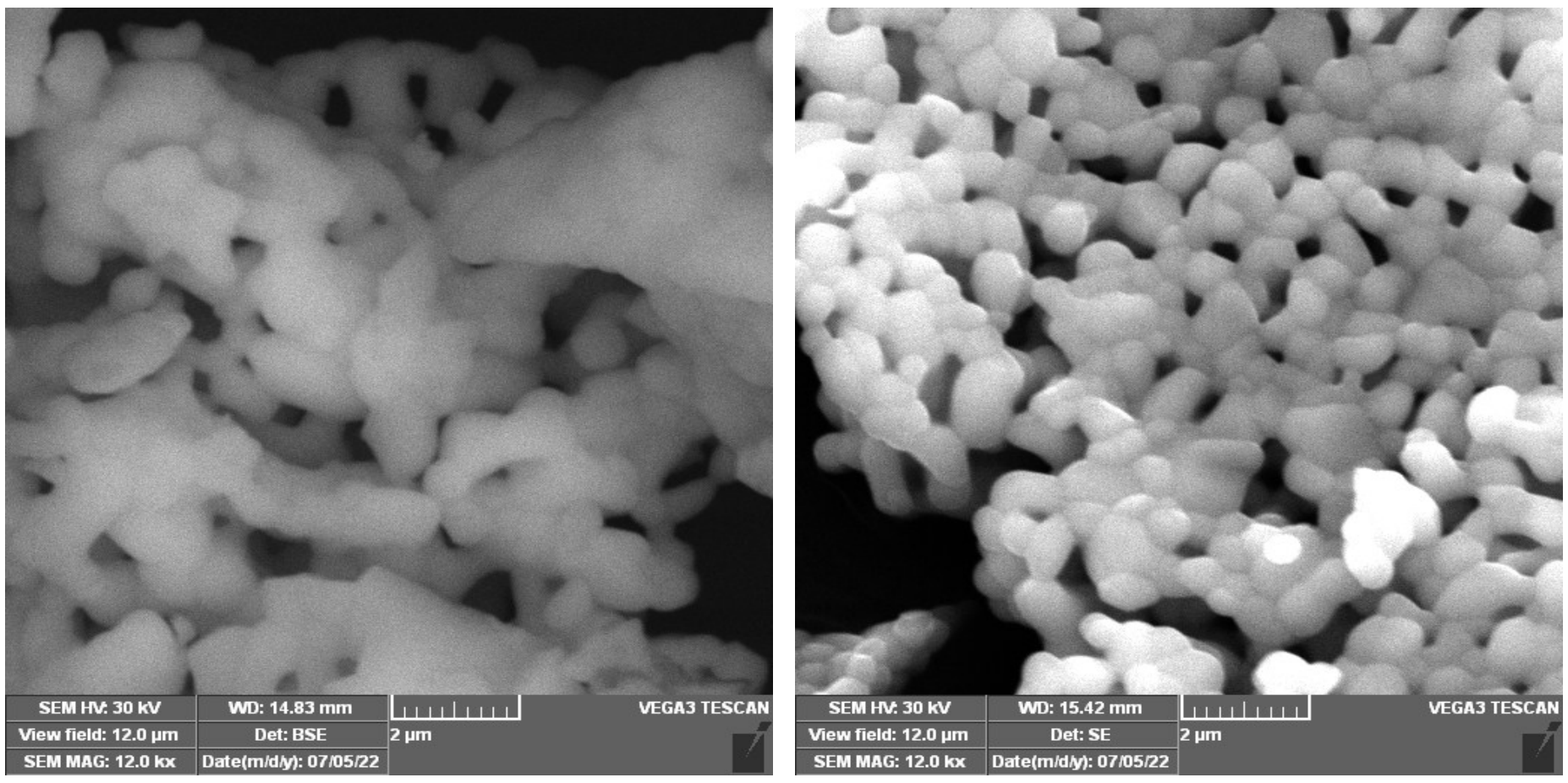

**Figure 10.** SEM for $BaY_2ZnO_5$ (left) BYZ 1-2 and (right) BYZ 1-4 pellet interiors.

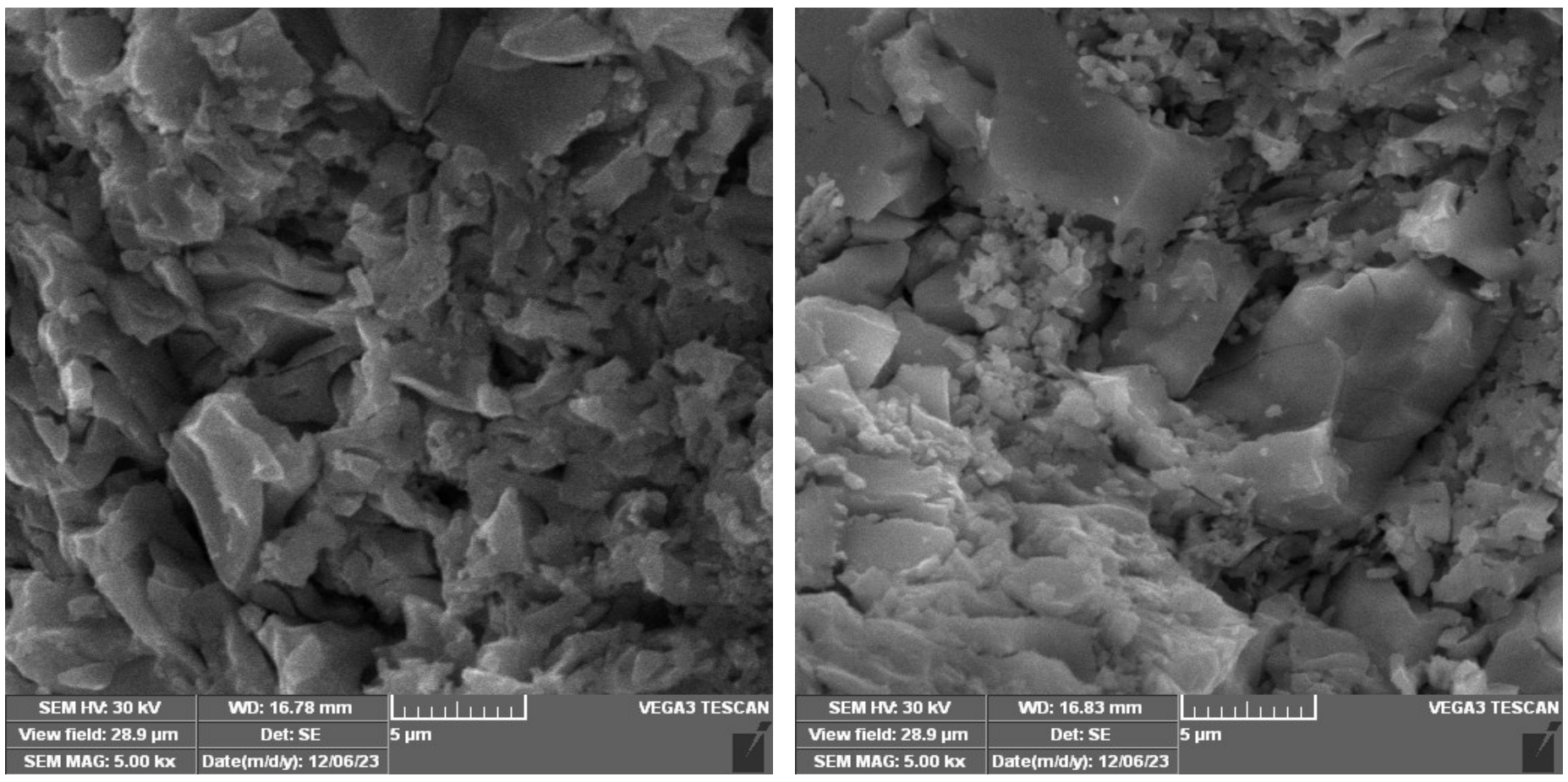

**Figure 11.** SEM for $Y_2TiO_5$ (left) 1-2 and (right) YTO 1-4 pellet interiors.

### *3.3. Upconversion spectra for the ceramics*

For obtaining upconversion photoluminescent spectra of the $Er^{3+}$, $Yb^{3+}$ doped oxidic matrix, the illumination power and the spectra acquisition times were adjusted for each type of ceramic due to large differences between the emission intensities, with BLZ having the smallest efficiency and BYZ having the greatest.

The acquisition time was chosen to be the smallest, in each case, in order to avoid the saturation of the sensor of the spectrometer in the visible domain, yet the useful signal to be maximal and all the spectra peaks to be visible and distinct.

The incident illumination power of the laser was set to 100 mW, and the acquisition times varied from 7 ms in the case of BYZ to 100 ms of YTO. From the relative values of these values, the relative efficiency can be inferred.

**Table 2** presents the results with YO taken as reference and the others transformed as for the acquisition time would be 25 ms and the illumination power would be 100 mW.

**Table 2**. Relative total emissions for the 1-4 cases normalized to YO 1-4, it can be seen that BYZ is 3.93 times more efficient than YO whereas YTO is three times less efficient.

| Sample | Illum Power (mW) | Acquisition time (ms) | Red + Green (total counts measured) | Intensity transformed (counts) | Relative intensity to YO 1-4. |
|---|---|---|---|---|---|
| YO 1-4 | 100 | 25 | 243232 | 243232 | 1.00 |
| BGZ 1-4 | 100 | 25 | 232198 | 232198 | 0.95 |
| BLZ 1-4 | 50 | 200 | 358405 | 89601 | 0.37 |
| BYZ 1-4 | 100 | 7 | 267350 | 954823. | 3.93 |
| YTO 1-4 | 100 | 100 | 339662 | 84915 | 0.35 |

When looking at the spectra in Figures 12-16, it should be noted the strong signal in the IR region of 950-1050 nm, which shows that the spectrometer was saturated on this zone.

The saturation is caused both by the reflection of the incident 975.5 nm light and by reemission by the $Yb^{3+}$ ions and shows how much the probe dissipates the incident energy instead of upconverting it into higher energy photons.

This saturation zone is very narrow in the case of BYZ, which has, as seen in **Table 2**, high efficiency, and is very wide in the case of YTO, which has the lowest efficiency.

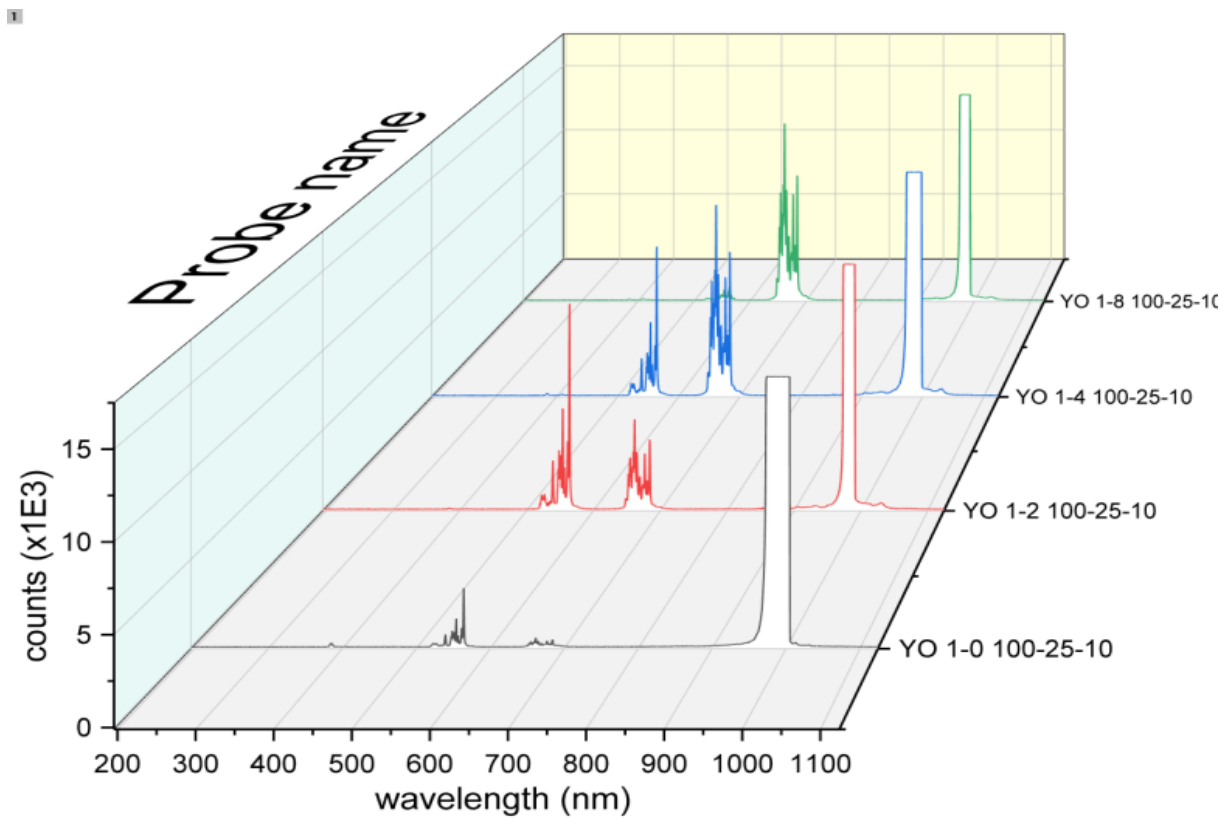

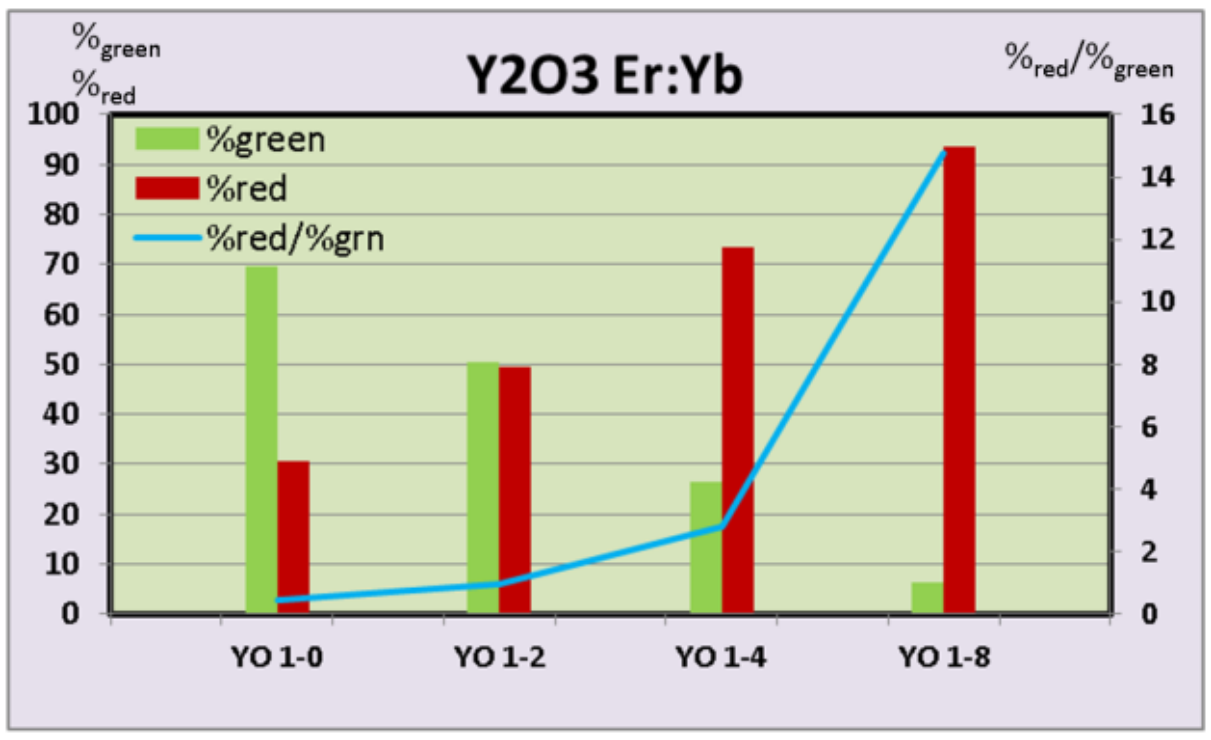

(A) (B)

**Figure 12.** (A): Spectra for $Y_2O_3$ (YO), concentrations 1-0, 1-2, 1-4, 1-8. (B): Percent of green (range of 515-580 nm) emission intensity vs red (range of 640-700 nm) emission intensity; Observe how the increase in $Yb^{3+}$ concentration increases the red emission. This is an indication that $Yb^{3+}$ promotes the populating of $^4F_{9/2}$ level of $Er^{3+}$.

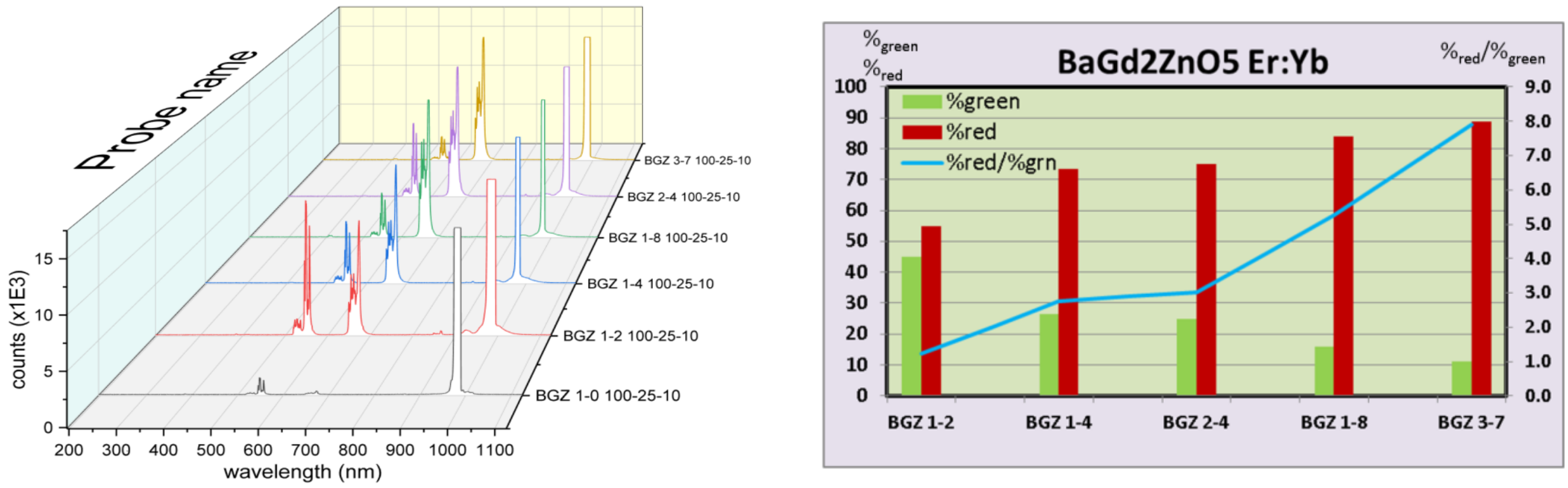

(A) (B)

**Figure 13.** (A): Spectra for $BaGd_2ZnO_5$ (BGZ), concentrations 1-0, 1-2, 1-4, 1-8, 2-4, 3-7. (B): Percent of green (range of 515-580 nm) emission intensity vs red (range of 640-700 nm) emission intensity. Observe how the increase in $Yb^{3+}$ concentration increases the red emission which is an indication that $Yb^{3+}$ promotes the populating of $^4F_{9/2}$ level of $Er^{3+}$.

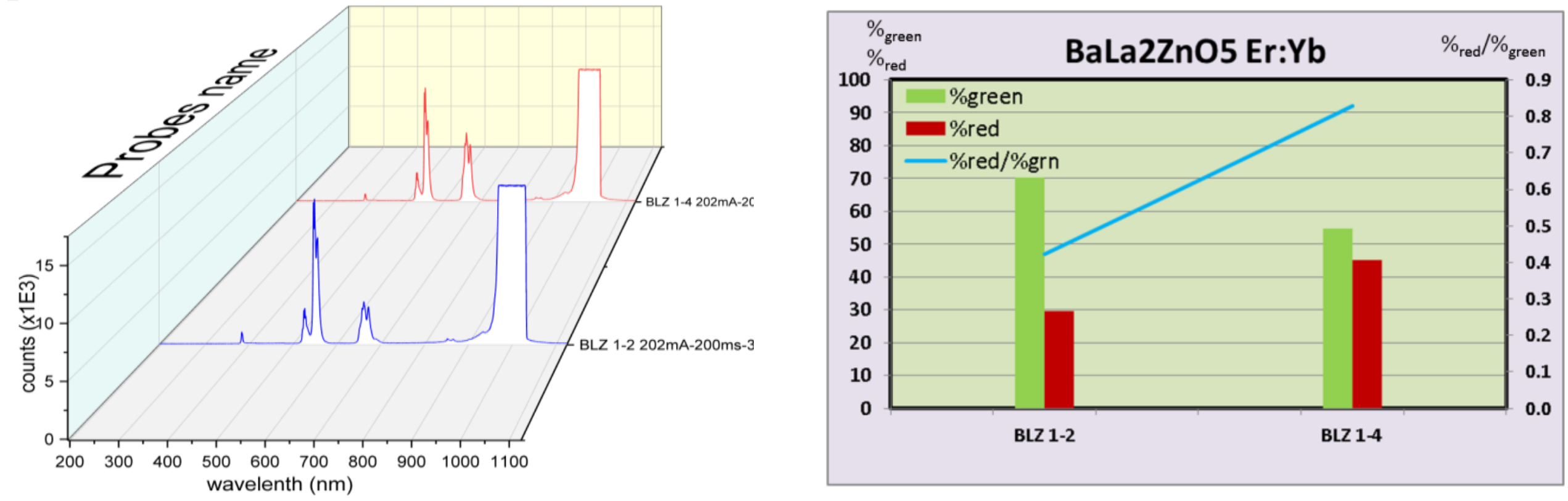

(A) (B)

**Figure 14.** (A): Spectra for $BaLa_2ZnO_5$ (BLZ), concentrations 1-2, 1-4,. (B): Percent of green (range of 515-580 nm) emission intensity vs red (range of 640-700 nm) emission intensity. Observe how the increase in $Yb^{3+}$ concentration determines the increase of the red emission indicating that $Yb^{3+}$ promotes the populating of $^4F_{9/2}$ level of $Er^{3+}$.

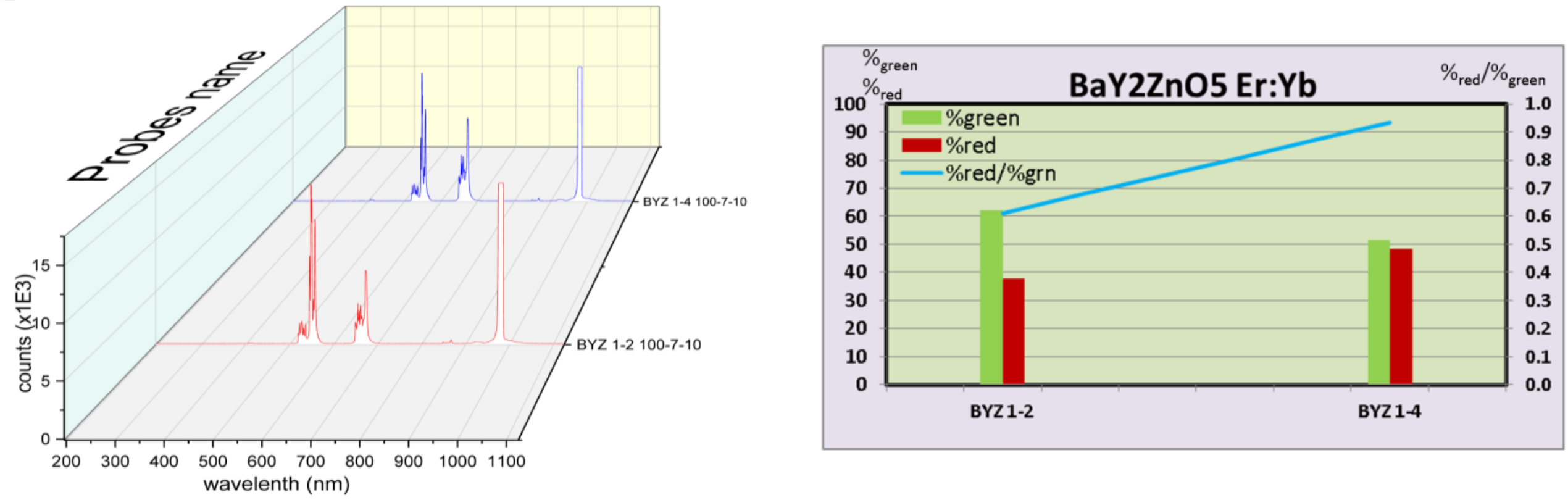

(A) (B)

**Figure 15.** (A): Spectra for $BaY_2ZnO_5$ (BYZ), concentrations 1-2, 1-4,. (B): Percent of green (range of 515-580 nm) emission intensity vs red (range of 640-700 nm) emission intensity. It can be seen that the increase in $Yb^{3+}$ concentration facilitates the red emission showing that $Yb^{3+}$ promotes the populating of $^4F_{9/2}$ level of $Er^{3+}$.

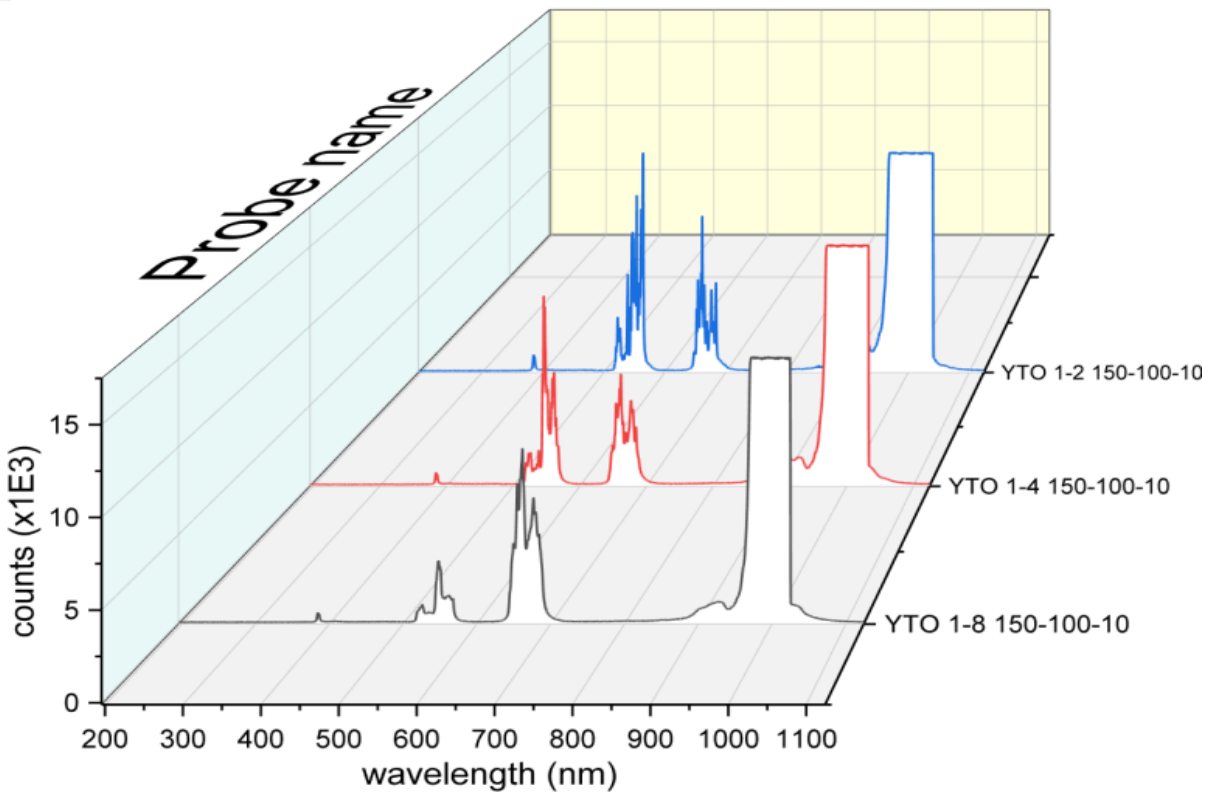

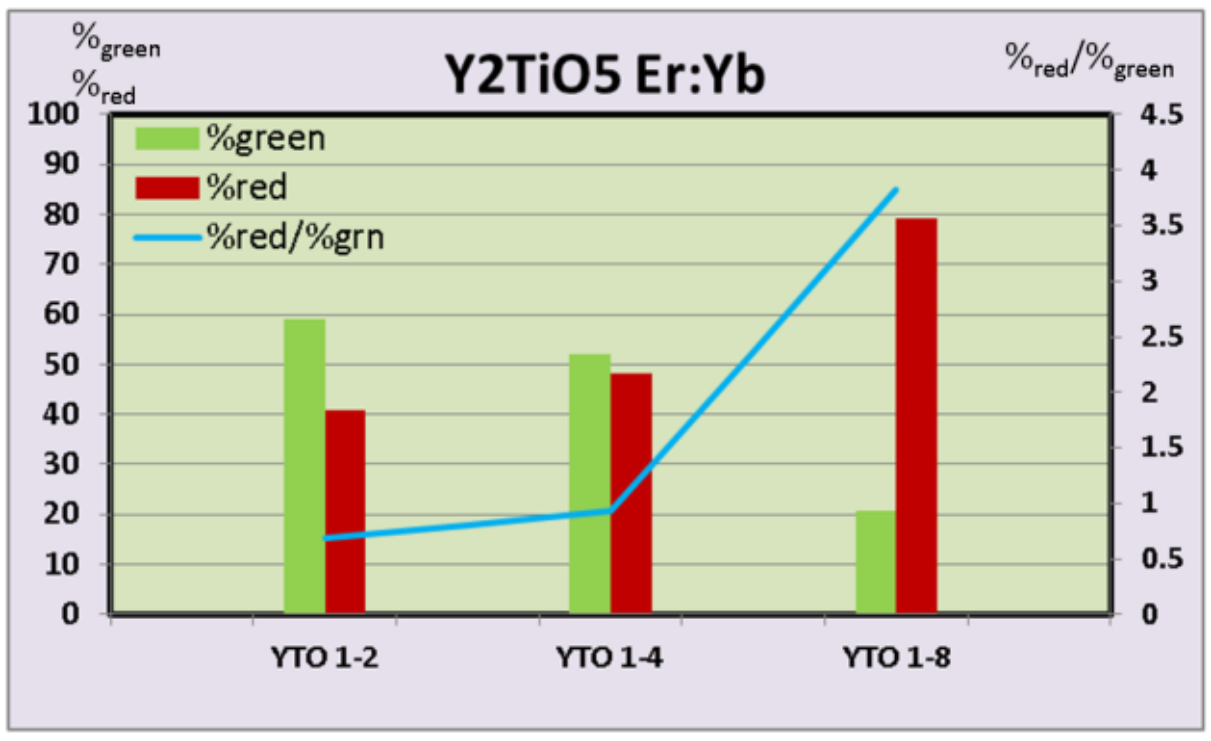

(A) (B)

**Figure 16.** (A): Spectra for $Y_2TiO_5$ (YTO), concentrations 1-2, 1-4, 1-8,. (B): Percent of green (range of 515-580 nm) emission intensity vs red (range of 640-700 nm) emission intensity. One can remark how the increase in $Yb^{3+}$ concentration determines the increase of the red emission indicating a link between $Yb^{3+}$ concentration and the population of $^4F_{9/2}$ level of $Er^{3+}$.

### *3.4. The relative intensities of the upconversion response.*

Figures 12-16 (B) are containing the graph showing the percent of red emission intensities (integral for range 640–700 nm) vs. green emission intensities (integral for range 515 nm–585 nm). The graphs reveal that the percent of red emission increases when the concentration of $Yb^{3+}$ is increased. This is an indication that $Yb^{3+}$ promotes the populating of 4F9/2 level of $Er^{3+}$. For the chosen range of concentrations, the red intensity variation (or green, respectively) is linearly correlated with the logarithm of $Yb^{3+}$ concentration, and this behavior is the same across all ceramic hosts, revealing subtleties of the energy transfers between $Er^{3+}$ and $Yb^{3+}$.

For this behavior, we didn't find in the literature any explanation, and this comparative study can act as a starting point for research in this direction.

### *3.5. Comparison for all emissions for Er1% Yb4% case*

**Figure 17**(A),(B),(C), presents the emissions for all cases with Er1% and Yb4%. They are shown in order to compare the peaks positions and the splittings generated by the crystal field of the hosts. The plots are scaled differently and offseted along y axis such to be easily compared

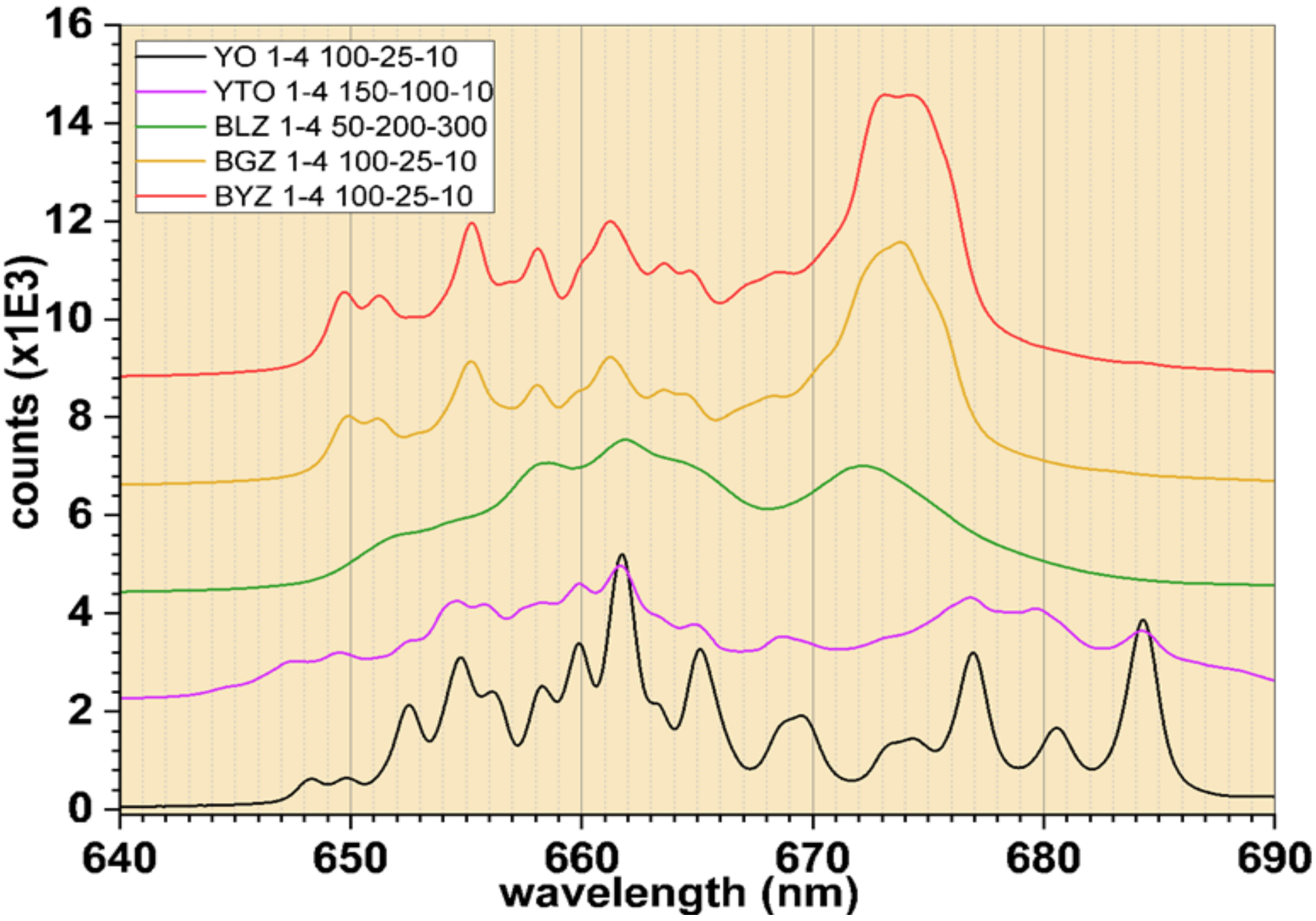

(A) $Er^{3+}$: $^4F_{9/2}\rightarrow{}^4I_{15/2}$

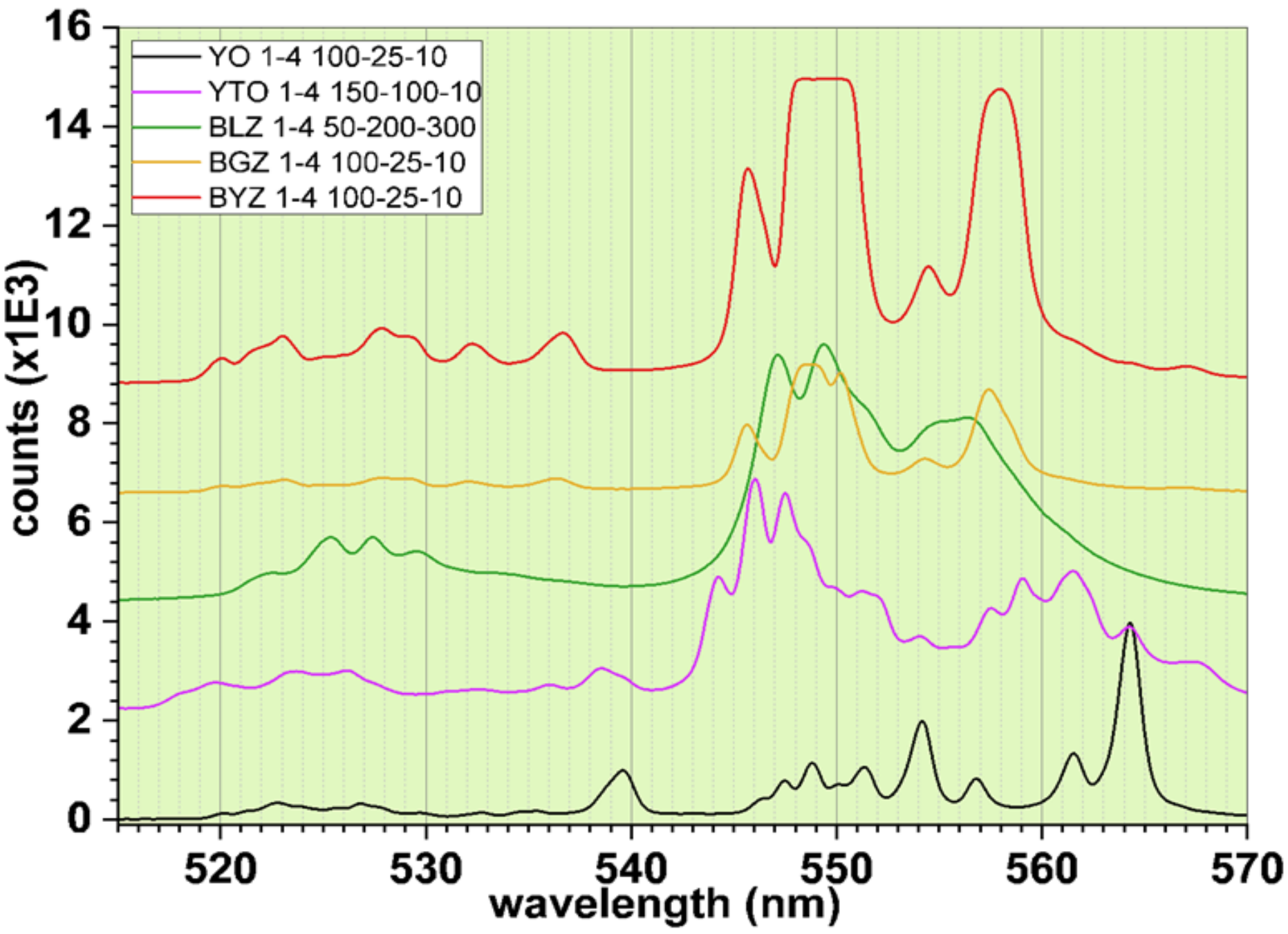

(B) $Er^{3+}$: $^2H_{11/2}\rightarrow{}^4I_{15/2}$ $Er^{3+}$: $^4S_{3/2}\rightarrow{}^4I_{15/2}$

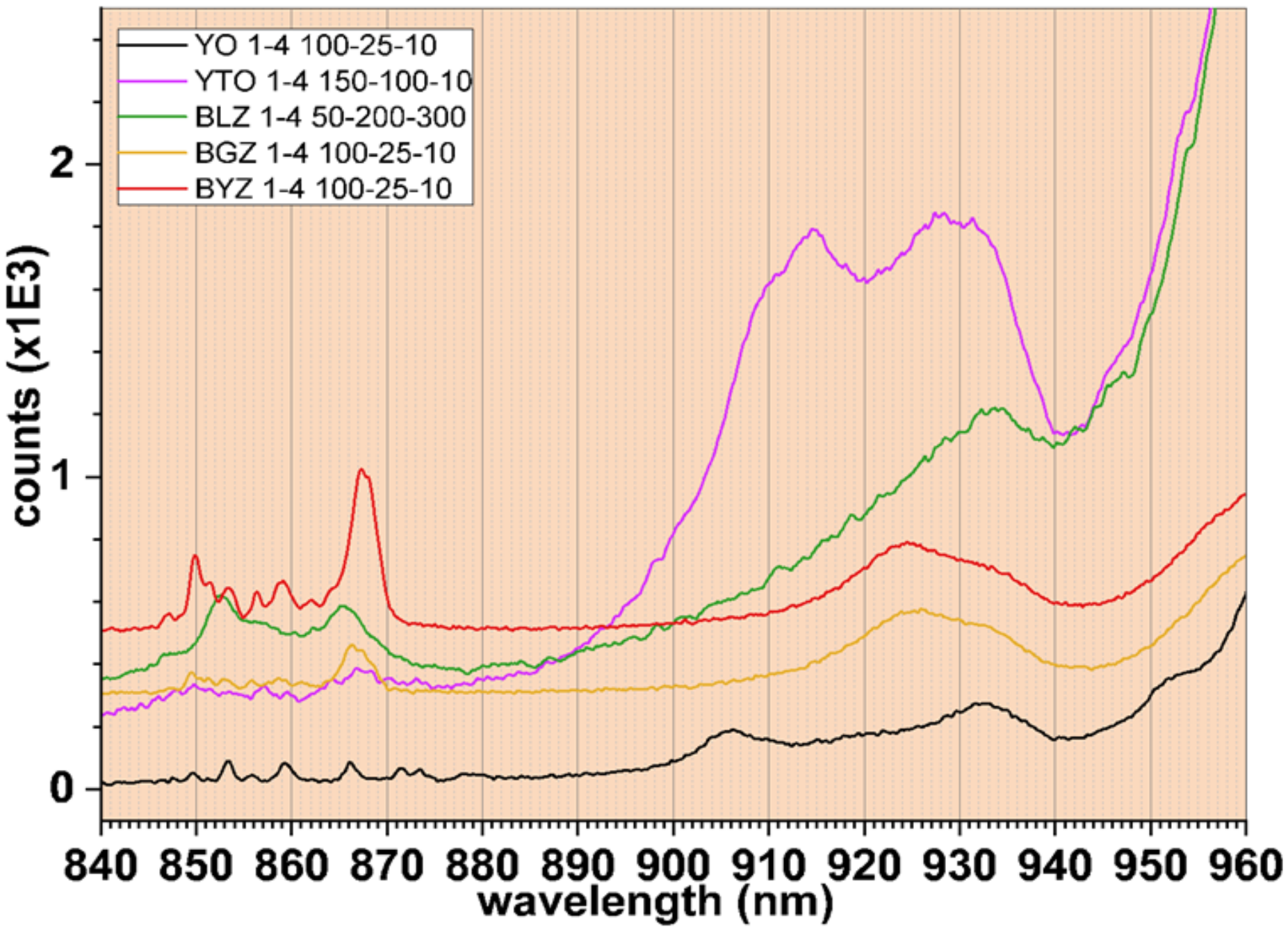

(C) $Er^{3+}$:$^4F_{9/2}\rightarrow{}^4I_{13/2}$ (840 - 885 nm) $Yb^{3+}$: $^2F_{5/2}\rightarrow{}^2F_{7/2}$ anti-Stokes band (900 - 940 nm)

**Figure 17**. All graphs are individually scaled such to be easily compared. Also, they are shifted on y axis. The plots are for each transition specified below each image. YO has the strongest crystal field of all cases distinctly resolving the energy levels, while BLZ has very broadened lines. Remark the phonon sidebands of $Yb^{3+}$ (900-940 nm), from which the phonon energies for each matrix can be inferred.

## 4. Discussion

As seen in the chapter with the comparative behavior of $Er^{3+}$ upconversion luminescence in the crystalline ceramic hosts we've studied, the more sensitizer $Yb^{3+}$ ions are present in the matrix, the more the red component of emission of $Er^{3+}$ is enhanced while the green emission is diminished.

Since this behavior is constant across all the crystalline hosts studied, regardless of their particular characteristics, like composition, density, crystal structure, symmetry of embedding site of $Er^{3+}$, dielectric constant, etc., it is clear that the parameter that governs this phenomenon is the medium distance between $Er^{3+}$ and $Yb^{3+}$ ions, whose value is, for most part at low concentrations, greater not only than the crystal's unit cell dimensions but even greater than the nanocrystal sizes that form the ceramic.

So, an evaluation of these average interionic distances, which depend on the concentrations of $Er^{3+}$ and $Yb^{3+}$, is mandatory for the understanding of the phenomenon of the aforementioned red shift. The effective measurement of these distances is impossible, at least for our possibilities, so the single way of assessing them, at least as order of magnitude, is the computer simulation.

This was done in [36], and the conclusion was that the $Yb^{3+}$ ions act as mirroring cavities surrounding $Er^{3+}$ ions, increasing the trapping efficiency of the incoming 980 nm radiation.

## 5. Conclusion

A comparison between different types of crystalline hosts with the same concentrations of $Er^{3+}$ and $Yb^{3+}$ ions was made. X-ray diffractograms of the probes, showing that the obtained structures were those expected, are also presented.

The upconversion spectra of the samples were shown for the cases Er:Yb of 1:2, 1:4, and 1:8, and the relative red/green percentual content in the total visible emission was plotted. For the 1:4 case, the spectra were shown in a comparative and very detailed way in order to observe the crystal field strengths and the individual peaks positions.

The comparison shows that percentual red vs. green emissions increase linearly with the increase of the logarithm of the relative Yb-Er concentration, a behavior that is constant in this range of concentrations across the kinds of crystalline hosts. This finding shows undoubtedly that the $Yb^{3+}$ ion sensitization effect on $Er^{3+}$ ions is a phenomenon determined by the interionic distance. The laws of this relationship will require further research.

The intensity of the emissions for a constant illumination were also tabulated and compared. The BYZ case was the most efficient, having an intensity almost 4 times higher than YO, which was used as a reference, while BLZ was on par with YTO, having efficiencies only a third of YO.

**Author Contributions:** Conceptualization, L.D.; methodology, D.B., C.M., and L.D.; software, L.D.; validation, L.D., D.B. and C.M.; formal analysis, L.D.; investigation, L.D.; resources, D.B., C.M. and L.D.; data curation, L.D.; writing—original draft preparation, L.D.; writing—review and editing, L.D., D.B. and C.M.; visualization, L.D. and C.M.; supervision, L.D., D.B. and C.M. All authors have read and agreed to the published version of the manuscript.

EOF